\documentclass[twocolumn,prl,nopacs,amsmath,amssymb]{revtex4}

\usepackage{graphicx}
\usepackage{color}
\usepackage{dcolumn}
\usepackage{bm}
\include{epsf}

\begin{document}

\title{Controllable Optical Negative Refraction and Phase Conjugation in Graphene}

\author{Hayk Harutyunyan$^1$, Ryan Beams$^1$, and Lukas Novotny$^{1,2}$}
\affiliation{$^{1}$ Institute of Optics, University of Rochester, Rochester, NY 14627, USA}
\affiliation{$^{2}$ ETH Z{\"u}rich, Photonics Laboratory, 8093 Z{\"u}rich, Switzerland}

\date{\today}


\maketitle

\noindent The development of optical metamaterials has resulted in the demonstration of remarkable physical properties, including cloaking, optical magnetism, and negative refraction~\cite{schurig06,Smith2004,soukoulis11}. The latter has attracted particular interest, mainly because of its promise for super-resolution imaging~\cite{Pendry2000,Fang2005,taubner06}. In recent years, negative refraction has been demonstrated with plasmonic materials~\cite{lezec07,Valentine2008,Palomba2012}
and nonlinear discrete elements~\cite{Katko2010}.
However,  the widespread use of negative refraction at {\em optical} frequencies is limited by  high losses  and strong dispersion effects, which typically limits operation to narrow frequency bands~\cite{stockman07b}. Here we use degenerate four-wave mixing (d-4WM) to demonstrate controllable negative refraction at a graphene interface, which acts as a highly efficient phase-conjugating surface. The scheme has very low loss because of the very small thickness of the nonlinear material 
($\sim 0.008\lambda$) 
and it ensures broadband operation due to the  linear bandstructure of graphene. \\[-2ex]

Phase-conjugation, has been used in optics for many decades to recover a beam profile that was distorted from passing through various materials~\cite{boyd02}. This is typically accomplished using d-4WM, in which two counter-propagating beams are used to write a holographic grating into a nonlinear crystal, while a third beam is diffracted in a direction opposite to the incident wave. The phase of the diffracted beam is reversed, which is equivalent to time-reversed propagation~\cite{boyd02, Pendry2008}. In order to be used for super-resolution imaging, two conditions have to be met. First, the nonlinear material must be much thinner than the wavelength so that it can be placed in close proximity to the object. Second, the phase-conjugate beam needs to be negatively refracted, propagating toward the detector instead of propagating backwards, in the opposite direction of the incident beam. By using a pair of closely spaced films of nonlinear material, it has been predicted that optical super-resolution can be achieved~\cite{Pendry2008}.\\[-2ex]

Here we demonstrate negative refraction using the nonlinear optical properties of  graphe\-ne~\cite{hendry10}. 
As shown in Fig.~\ref{fig1}(a), a negatively refracting lens is generated by pumping the graphene sample with two counter-propagating  beams ${\bf E}_1 \exp\!{(i\textbf{k}_1\cdot\textbf{r})}$ and ${\bf E}_2 \exp\!{(i\textbf{k}_2\cdot\textbf{r})}$ of wavelength $\lambda$.  A signal beam ${\bf E}_3 \exp\!{(i\textbf{k}_3\cdot\textbf{r})}$ of the same wavelength is then refracted at the holographic grating written by the two pump beams. 
%
%

\begin{widetext}
\begin{center}
\begin{figure}[!b]
	\includegraphics[width=41.0em]{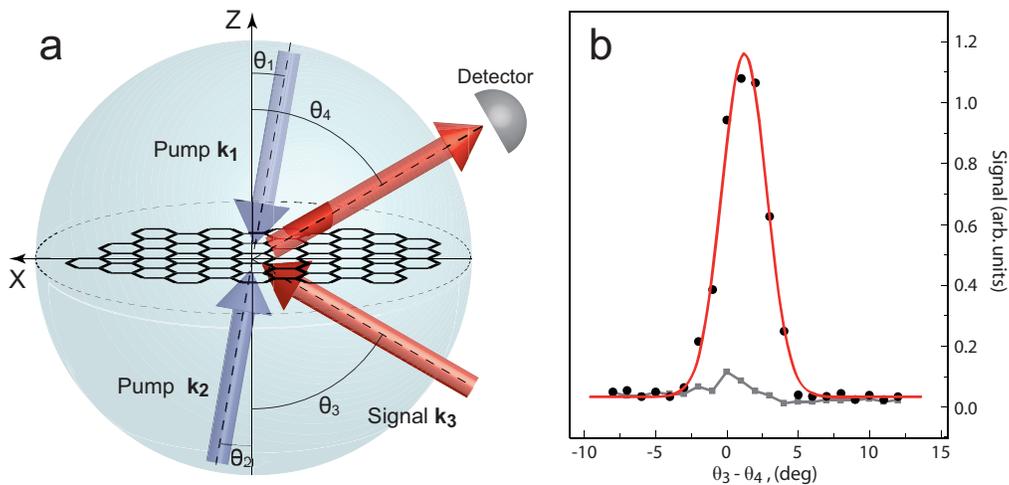}
	\vspace{-1em}
	\caption{({\bf a}) Illustration of the experiment. Two pump beams (blue arrows) are focused on a graphene flake (honeycomb lattice) at angles $\theta_1 = \theta_2$. The signal beam (lower red arrow) is refracted in the negative direction (upper red arrow).
The graphene sample is sandwiched between two glass hemispheres. ({\bf b}) Experimental verification. The intensity of the refracted light exhibits a peak at $\theta_4\approx\theta_3$, in agreement with negative refraction. The lower curve is a control measurement obtained by time-delaying the pulses of one of the pump beams.
\label{fig1}}
\end{figure}
\end{center}
\end{widetext}
The direction of the refracted beam follows from momentum conservation defined by the d-4WM process
\begin{equation}
{k}_{4,x}\;=\; {k}_{1,x}+{k}_{2,x}-{k}_{3,x} \; ,
\end{equation}
where ${k}_{i,x}=\textbf{k}_{i} \cdot \hat{\bf x}$, is the projection of the wavevector  $\textbf{k}_i$ on the graphene surface.
For counter-propagating pump beams we have $\textbf{k}_1+\textbf{k}_2=0$ and hence ${k}_{4,x}\;=\; -{k}_{3,x}$, consistent with negative refraction.
Thus, the two-dimensional graphene interface yields not only a phase-conjugate reflected wave that propagates against the incident signal beam,~\cite{boyd02} but also a negatively refracted beam, which is a prerequisite for realizing the super-resolution imaging scheme proposed in Ref.~\cite{Pendry2008}.\\[-2ex]

%

Our measurement scheme is  depicted in Fig.~\ref{fig1}(a). The output of a mode-locked Ti-Sapphire laser with center wavelength 800 nm and pulse duration of 200 fs is split into three beams, each of which is focused on the sample from a different angle of incidence. The focal spots of the three beams are overlapped on a graphene flake that is sandwiched between two glass hemispheres. 
A multimode fiber picks up the refracted light and sends it to a photodetector. 
%
%
The sample is placed on a 
rotation stage to tune the angles of incidence ($\theta_i$)  of the excitation beams relative to the normal of the graphene interface. The
two hemispheres guarantee that  sample rotation does not affect the alignment of the excitation beams.\\[-2ex] 

Fig.~\ref{fig1}(b) shows our experimental results. The refracted intensity features a resonance at $\theta_3-\theta_4 \approx 0^{\circ}$, that is,  the signal beam is refracted in negative direction as predicted by the momentum conservation condition. For $10\,$ mW of laser power in each of the excitation beams, the efficiency of negative refraction is found to be $0.1 \%$, which is remarkably high for a nonlinear process. The efficiency can be increased with higher laser powers and tighter focusing, approaching unity for realistic power density levels.\\[-2ex]

Our experimental results are the first demonstration of optical negative refraction based on phase-conjugation and time-reversal. The scheme provides all-optical control of negative refraction, which can be exploited for many optoelectronic applications.  Using graphene as an interface material ensures low-loss operation and virtually unlimited bandwidth. \\

This research was funded in part by the U.S. Department of Energy (grant DE-FG02-05ER46207). We thank John Pendry and Bradley Deutsch for valuable input and discussions.\\

Correspondence and requests for materials should be addressed to L.N.~(email:  lnovotny@ethz.ch) 


\end{document}